\documentclass[12pt]{article}
\usepackage{scicite}
\usepackage{xcolor}
\usepackage{times}
\usepackage{mathptmx}
\usepackage{graphicx}
\usepackage{hyperref}
\usepackage[T1]{fontenc}
\usepackage[utf8]{inputenc}
\usepackage{lipsum}
\usepackage{enumitem}
\usepackage{xcolor}

\definecolor{magenta}{rgb}{1.0, 0.0, 1.0}

\newcommand{\daa}{\Delta \alpha/\alpha}

\title{Four direct measurements of the fine-structure constant 13 billion years ago}

\author
{
Michael R. Wilczynska$^{1}$, 
John K. Webb$^{1\dagger}$, 
Matthew Bainbridge$^{2}$,\\
Sarah E. I. Bosman$^{3}$, 
John D. Barrow$^{4}$, 
Robert F. Carswell$^{5}$,\\
Mariusz P. D\k{a}browski$^{6}$, 
Vincent Dumont$^{7}$, 
Ana Catarina Leite$^{8,9,10}$,\\
Chung-Chi Lee$^4$, 
Katarzyna Leszczy\'nska$^{6}$, 
Jochen Liske$^{11}$,\\
Konrad Marosek$^{15}$, 
Carlos J.A.P. Martins$^{8,9}$, 
Dinko Milakovi\'c$^{12,13}$,\\
Paolo Molaro$^{14}$, 
Luca Pasquini$^{10}$\\\\
\small{$^{1}$University of New South Wales Sydney, Sydney NSW 2052, Australia}\\
\small{$^{2}$College of Science and Engineering, University of Leicester, University Road,}\\
\small{Leicester, LE17RH, UK}\\
\small{$^{3}$Department of Physics and Astronomy, University College London, WC1E 6BT, UK}\\
\small{$^{4}$DAMTP, Centre for Mathematical Sciences, University of Cambridge, Cambridge CB3 0WA, UK}\\
\small{$^{5}$Institute of Astronomy, Madingley Road, Cambridge CB3 0HA, UK}\\
\small{$^{6}$Institute of Physics, University of Szczecin, Wielkopolska 15, 70-451 Szczecin, Poland}\\
\small{$^{7}$ Lawrence Berkeley National Laboratory, Berkeley, CA, USA}\\
\small{$^{8}${Centro de Astrof\'{\i}sica da Universidade do Porto, Rua das Estrelas, 4150-762 Porto, Portugal}}\\
\small{$^{9}$Instituto de Astrof\'{\i}sica e Ci\^encias do Espa\c co, CAUP,
Rua das Estrelas, 4150-762 Porto, Portugal,}\\
\small{$^{10}$Faculdade de Ci\^encias, Universidade do Porto, Rua do Campo Alegre, 4150-007 Porto, Portugal}\\
\small{$^{11}$Hamburger Sternwarte, Universit{\"a}t Hamburg, Gojenbergsweg 112,}\\
\small{D-21029 Hamburg, Germany}\\
\small{$^{12}$European Southern Observatory, 85748 Garching bei München Germany}\\
\small{$^{13}$Ludwig-Maximilians-Universität, 80799 Munich, Germany}\\
\small{$^{14}$National Institute for Astrophysics, Astronomical Observatory of Trieste,}\\
\small{Via G.B. Tiepolo 11, I34134 Italy}\\
\small{$^{15}${Maritime University, Wały Chrobrego 1-2, 70-500 Szczecin, Poland}}\\
\\
\normalsize{$^\dagger$jkw@phys.unsw.edu.au}
}

\date{}

\begin{document} 
\maketitle 
\clearpage

{\bf Abstract: Observations of the redshift $z=7.085$ quasar J1120+0641 have been used to search for variations of the fine structure constant, $\alpha$, over the redshift range $5.5$ to $7.1$. Observations at $z=7.1$ probe the physics of the universe when it was only 0.8 billion years old. These are the most distant direct measurements of $\alpha$ to date and the first measurements made with a near-IR spectrograph. A new AI analysis method has been employed. Four measurements from the {\sc x-shooter} spectrograph on the European Southern Observatory's Very Large Telescope (VLT) directly constrain any changes in $\alpha$ relative to the value measured on Earth ($\alpha_0$).  The weighted mean strength of the electromagnetic force over this redshift range in this location in the universe is $\daa = (\alpha_z - \alpha_0)/\alpha_0 = (-2.18 \pm 7.27) \times 10^{-5}$, i.e. we find no evidence for a temporal change from the 4 new very high redshift measurements. When the 4 new measurements are combined with a large existing sample of lower redshift measurements, a new limit on possible spatial variation of $\daa$ is marginally preferred over a no-variation model at the $3.7$$\sigma$ level.}

\section*{Main text}

What fundamental aspects of the universe give rise to the laws of Nature? Are the laws finely-tuned from the outset, immutable in time and space, or do they vary in space or time such that our local patch of the universe is particularly suited to our existence? We characterize the laws of Nature using the numerical values of the fundamental constants, for which increasingly precise and ever--distant measurements are accessible using quasar absorption spectra.

The quest to determine whether the bare fine structure constant, $\alpha$, is indeed a constant in space and time has received impetus from the recognition that the possibility that there are additional dimensions of space, or that our constants are partly or wholly determined by symmetry breaking at ultra-high energies in the very early universe. The first proposals for time variation in $\alpha$ by Stanykovich \cite{Stan}, Teller \cite{Tell} and Gamow \cite{Gam} were actually motivated by the large numbers coincidences noted by Dirac \cite{dir, BT} but were quickly ruled out by observations \cite{dyson}. This has led to an extensive literature on varying constants that is reviewed in refs. \cite{uzan, dam, land, ol, Martins2017}.

There are also interesting new problems that have been about extreme fine tuning of quantum corrections in theories with variation of $\alpha$ by O'Donoghue \cite{Odon} and Marsh \cite{mar}. Accordingly, self-consistent theories of gravity and electromagnetism which incorporate the fine structure `constant' as a self-gravitating scalar field with self-consistent dynamics that couple to the geometry of spacetime, have been formulated in refs. \cite{bek, poly, SBM, BLip1, BLip2, BLip3, BLip4} and extended to the Weinberg-Salam theory in refs. \cite{EW1, EW2}. They generalise Maxwell's equations and general relativity in the way that Jordan-Brans-Dicke gravity theory \cite{jor, BD} extends general relativity to include space or time variations of the Newtonian gravitational constant, $G$, by upgrading it to become a scalar field. This enables different constraints on a changing $\alpha (z)$ at different redshifts, $z$, to be coordinated; it supersedes the traditional approach \cite{dyson2} to constraining varying $\alpha$ by simply allowing $\alpha$ to become a variable in the physical laws for constant $\alpha$. Further discussions relating spatial variations of $\alpha$ to inhomogeneous cosmological models can be found in \cite{Dab14,Balc17}.

Direct measurements of $\alpha$ are also important for testing dynamical dark energy models, since they help to constrain the dynamics of the underlying scalar field \cite{Martins2017} and thus dynamics can be constrained (through $\alpha$) even at epochs where dark energy is still not dominating the universe. Indeed, the possibility of doing these measurements deep into the matter era is particularly useful, since most other cosmological datasets (coming from type Ia supernovas, galaxy clustering, etc) are limited to lower redshifts.

The inputs needed for these theories come from a variety of different types of astronomical observations: high-precision observations of the instantaneous value of $\alpha (z_{i})$ characterising quasar spectra at various redshifts $z_{i}$ to test possible time variations; both local and non-local measurements of $\alpha (\vec{x})$ at different positions in the universe \cite{1999PhRvL..82..884W, 2003MNRAS.345..609M, King12} to search for spatial variation; the CMB \cite{jdb1, 1999PhRvD..60b3516K}; the Oklo natural reactor \cite{oklo1, oklo2, oklo3, oklo4}; atomic clocks \cite{uzan, flam, 1999PhRvL..82..888D}; compact objects in which the local gravitational potential may be different \cite{MBS, VVF} or atomic line separations in white dwarf atmospheres \cite{WD1}.

As a result of these observational searches for evidence of varying $\alpha$, there has been a persistent signal of spatial variation at a level of $\sim 4\sigma$ from detailed studies of large numbers of quasar spectra \cite{Webb2011, mp2017, DW17}, motivating further direct measurements, especially by extending the measurement redshift range. The relative wavelengths of absorption lines imprinted on spectra of background quasars are sensitive to the fine-structure constant, $\alpha = e^{2}/\hbar c$ (where $e$, $\hbar$, and $c$ are the electron charge, the reduced Planck's constant, and the speed of light). Comparing quasar measurements with high precision terrestrial experiments provides stringent constraints on any possible spacetime variations of the fine-structure constant, as predicted by some theoretical models, \cite{Sandvik2002, Uzan2011, Stadnik2015, vandeBruck2015, Martins2017}. The quasar J1120+0641 \cite{2011Natur.474..616M} is of particular interest in this context because of its very high redshift.  Its emission redshift is $z=7.085$, corresponding to a look-back time of 12.96 billion years in standard $\Lambda$CDM cosmology. J1120+0641 is one of the most luminous quasars known \cite{Barnett2015}, enabling high spectral resolution at high signal to noise. We make use of spectra obtained using the {\sc x-shooter} spectrograph \cite{Vern11} on the European Southern Observatory's Very Large Telescope (VLT), with nominal spectral resolution $R = \lambda/\Delta\lambda = 7000 - 10,000$ \cite{Bos17}. The total integration time is 30 hours. Data reduction, continuum fitting, and absorption system identification are discussed in \cite{Supp_this_paper}.

The {\sc x-shooter} instrument provides a broad spectral wavelength coverage. This maximises the discovery probability of absorption systems along the sightline, enabling the identification of potential coincidences (i.e. blends) between absorption species at different redshifts, an essential step in making a reliable measurement of $\alpha$. In all, 11 absorption systems are detected \cite{Bos17,Supp_this_paper}. Desirable characteristics of an absorption system are a selection of transitions with different sensitivities to a change in $\alpha$ and a velocity structure in the absorbing medium that is as simple as possible.  

Of the 11 absorption systems identified along the J1120+0641 sightline (Table \ref{tab:identifications}), four are found to be suitable for a measurements of $\alpha$, at redshifts $z_{abs} = 7.059, 6.171, 5.951$, and $5.507$. The atomic transitions used to measure $\alpha$ in these four systems are highlighted in (Table \ref{tab:identifications}). The highest redshift system has, of the four, the least sensitivity to varying $\alpha$. No other direct quasar absorption $\alpha$ measurements have previously been made at such high redshift. Prior to the measurements described in this paper, the highest redshift quasar absorption direct measurement of $\alpha$ was at $z=4.1798$ \cite{Murphy2004}. Voigt profile models for each of the four absorption systems were automatically constructed using a genetic algorithm, {\sc gvpfit}, which requires no human decision making beyond initial set-up parameters \cite{gvpfit17}. The genetic part of the procedure controls the evolution of the model development. {\sc vpfit} \cite{vpfit} is called multiple times within each generation to refine the model which then becomes the parent for subsequent generations. Absorption model complexity increases with each generation. A description of {\sc gvpfit} can be found in \cite{gvpfit17} and an assessment of its performance in \cite{Bainbridge2017}. The procedure out-performs human interactive methods in that it gives objective, reproducible, and robust results, and introduces no additional systematic uncertainties. The method is computationally demanding, requiring supercomputers.  New procedures have been introduced for the analysis in this paper, beyond those described in \cite{gvpfit17}, so are described here.

The analysis of each of the four absorption systems took place in 4 stages. Throughout, $\daa$ is kept as a free fitting parameter, making use of the {\it Many Multiplet Method} \cite{Dzuba99,Webb99}. In Stage 1 we imposed the requirement that all velocity components are present in all species being fitted, irrespective of line strength. Without this requirement, an absorbing component in one species might fall below the detection threshold determined by the spectra data quality, but not in another. This requirement was only applied in this first Stage because it was found in practice to help model stability by discouraging the fitting procedure from finding a model with implausibly large $b$ or high $N$ in one or more components.  The requirement is dropped subsequently. {\sc gvpfit} was allowed to evolve (that is, the complexity of the model was allowed to increase) for the number of generations required to pass through a minimum value of the corrected Akaike Information Criterion statistic (AICc) \cite{Aka74,Sug78}.  The model resulting from this first Stage of the analysis is the model at which AICc is at a minimum and is already quite good but is not final.

In Stage 2 we use the model from Stage 1 as the parent model input to {\sc gvpfit} but now drop the requirement that all velocity components are present. The other requirements from Stage 1 were carried over to Stage 2. At this Stage, one further increase in model complexity is introduced.  Although the spectral continuum model was derived before the line fitting process, we allow for residual uncertainties in continuum estimation where needed by including introducing additional free parameters allowing the local continuum for each region to vary using a simple linear correction as described in the {\sc vpfit} manual\footnote{\url{http://www.ast.cam.ac.uk/~rfc/vpfit11.1.pdf}}. The minimum AICc model from this stage is again taken as the parent model for the next Stage.

In Stage 3 we check to see whether any interloping absorption lines from other redshift systems may be present within any of the spectral regions used to measure $\alpha$. When interloper parameters are introduced, degeneracy can occur with other parameters associated with the metal lines used to measure $\alpha$. To avoid this problem, all previous parameters are temporarily fixed and {\sc gvpfit} is used in a first pass to identify places in the data where the current model is inadequate. Interlopers, modelled as unidentified atomic species, are added automatically by {\sc gvpfit} to improve the current fit.

In Stage 4, the model resulting from this third Stage is used as the input model for the fourth and final part of the process, which entails running {\sc gvpfit} again but this time with all parameters free to vary (subject to the physical constraint that all $b$-parameters are tied and all redshifts of corresponding absorbing components are tied, as was the case throughout all Stages).

In previous non-AI analyses, the general approach was to construct absorption system models based on turbulent broadening \cite{King12} and then to construct a thermal model {\it from} the turbulent parameters. One significant advantage of the AI approach is that it is straightforward to build turbulent and thermal models independently and this has been done for all four absorption systems reported here. Whilst this was possible prior to {\sc gvpfit} automation, it was very time consuming to do manually and therefore was not done. The Doppler, or {\it b}-parameters of different ionic species are related by $b^{2}_{i}=2kT/m_{i}+b^{2}_{turb}$ where the {\it i$^{th}$} ionic species has mass {\it m}, {\it k} is Boltzmann's constant and {\it T} is the temperature of the absorption cloud. The first term describes the thermal contribution to the broadening of the {\it b}-parameters and the second term the contribution of bulk, turbulent motions. If the line widths for a particular absorption cloud are dominated by thermal broadening, the second term of the equation is zero and vice-versa if the broadening mechanism is predominantly turbulent. These two cases are the limits of possibilities for the values of the {\it b}-parameters.

We have modelled each absorption system using the two limiting cases: first assuming the lines are thermally broadened and then assuming turbulent broadening. Modelling in this way results in two measurements of the fine-structure constant for each absorption system. Table \ref{tab:results} gives the results, which show that both measurements, for all four absorption systems, are consistent with each other.  Rather than discarding the highest $\chi^2$ model, since both models are statistically acceptable, we give a single value of $\alpha$ from our results using the method of moments estimator to determine the most likely value. The method of moments estimator compares the weighted relative goodness of fit differences between the thermal and turbulent models. This method is conservative, in that it only ever increases the uncertainty estimate on $\alpha$ from the smallest one and accounts for cases where the fits are consistent (like our results) and where inconsistent, the value chosen is more heavily weighted to the model with a lower $\chi^2$ (see \cite{King12}).

\begin{figure}
\centering
\includegraphics[width=16cm]{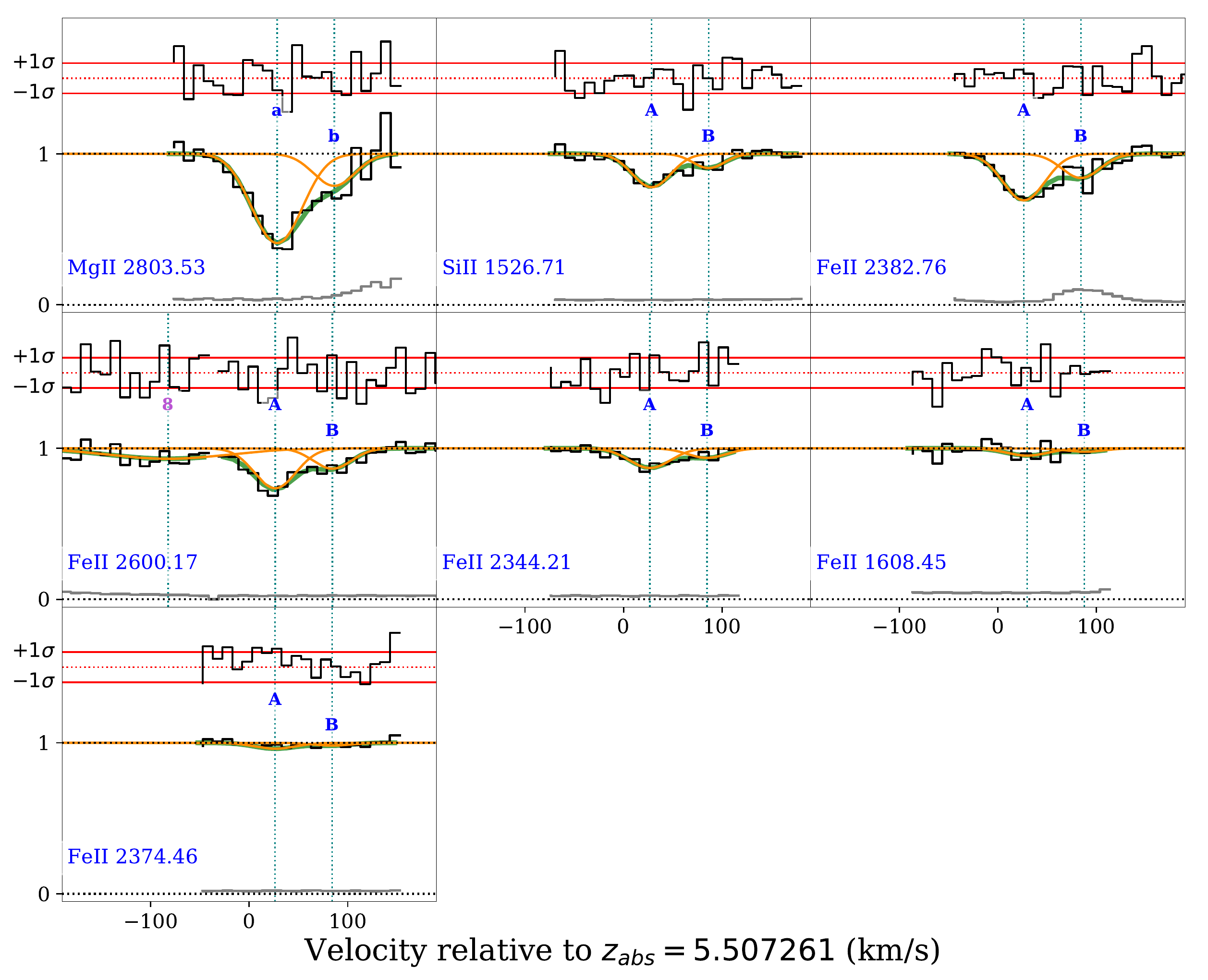}
\caption{Transitions used (black histogram) and absorption system model (thick green continuous line) from {\sc gvpfit} for the $z_{abs}=5.50726$ system. The model shown is the thermal fit. In all fits, model profiles included isotopic structures assuming relative terrestrial abundances for all species. Individual absorption components are illustrated by the thinner continuous orange lines. The grey line near the bottom of each panel shows the 1-$\sigma$ uncertainty on each spectral pixel. The upper black histogram illustrates the normalised residuals. The horizontal red dotted line is a reference line (arbitrarily offset for clarity) representing the expected mean of zero for the normalised residuals.  The horizontal red continuous lines illustrate the expected $\pm 1-\sigma$ deviations. Components labelled 'a' and 'b' indicate reference transitions and upper-case letters 'A' and 'B' indicate parameters tied to the reference transitions (\small{\url{http://www.ast.cam.ac.uk/~rfc/vpfit11.1.pdf}}). Where cosmic rays have fallen on the quasar spectrum, pixels have been clipped (as can be seen by the gaps in the black histograms). The MgII 2796 line fell in a region of the spectrum with an incompletely removed telluric line so was excluded from the fitting process. Plots for the other many-multiplet absorption systems observed in this spectrum are provided in the supplementary material.}
\label{fig:therm5-50}
\end{figure}

Figure \ref{fig:therm5-50} illustrates one model for the lowest redshift system analysed (see caption for details). All final model parameters associated with the four high redshift absorption systems modelled here are provided in the online Supplementary Materials, which also describes upper limits on potential systematic effects due to wavelength distortions. This is the first time multiple absorption systems along a given sightline have been simultaneously modelled in order to constrain the presence and impact of long-range wavelength distortions across a large wavelength range. This is important because in this way the distortion model parameters are more tightly constrained and hence the possible additional systematic error on $\daa$ is minimised. We find that in this case the additional systematic is smaller than the statistical uncertainty on $\daa$ from {\sc vpfit}.

\begin{figure}
\centering
\includegraphics[origin=c,width=\columnwidth]{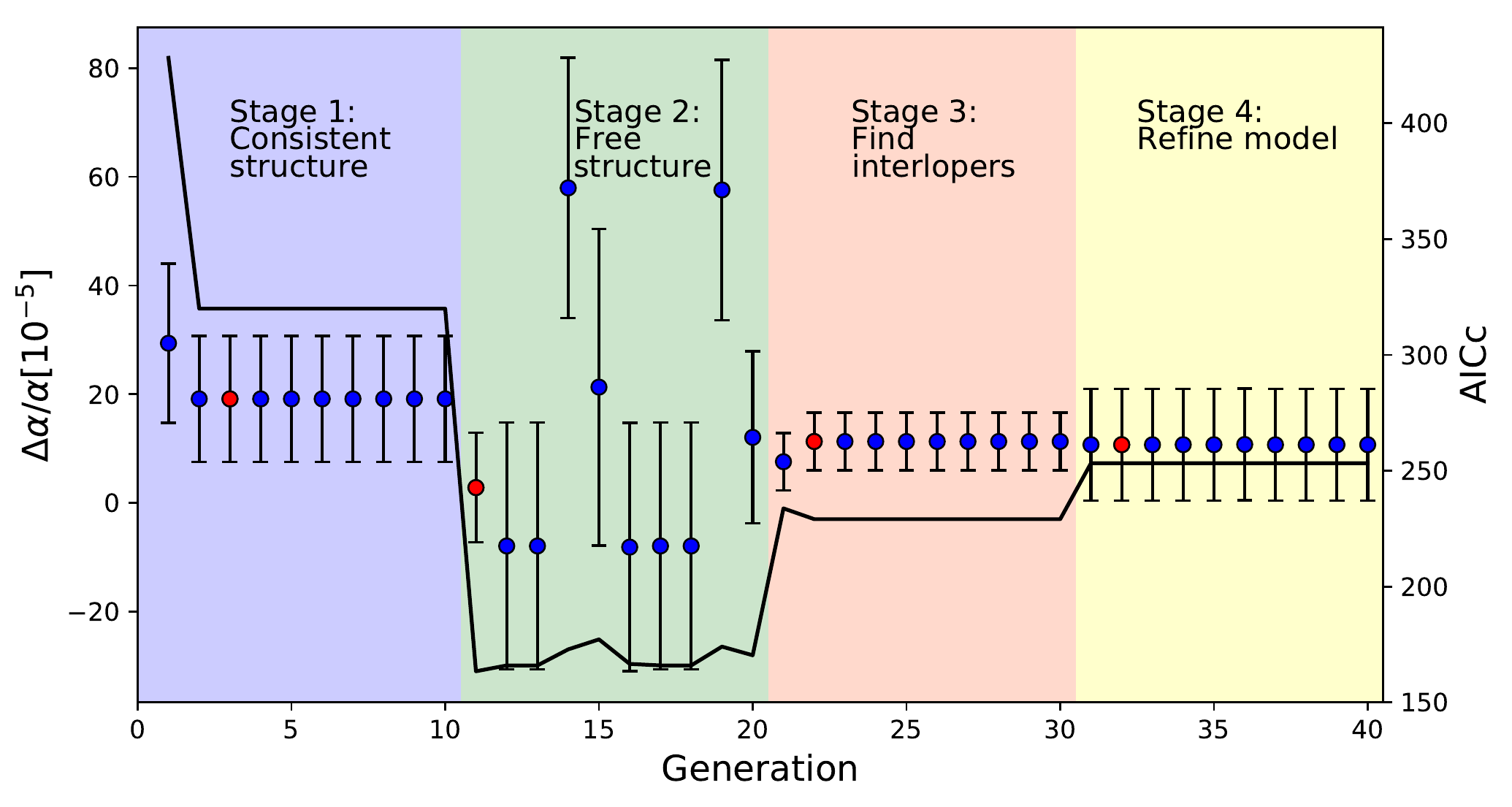}
\caption{Illustration of the {\sc gvpfit} procedures used in obtaining the $\daa$ measurement for the $z_{abs}=5.50726$ absorption system. The 4 fitting Stages are indicated by the 4 different shaded regions (Stage 1: Consistent structure required; Stage 2: Consistent structure requirement removed; Stage 3: Find interlopers; Stage 4: Final tied parameter fit.  See text for details). Each point illustrates the lowest $\chi^2$ point at each generation. Model complexity increases with generation number. The error bars are artificially small in Stage 3 because some parameters were fixed during the initial interloper fit. The red point in each Stage indicates the model with the smallest AICc for that Stage. The continuous black line illustrates the AICc. The final model for this system is indicated by the red point at generation 32 in Stage 4. The ``plateauing'' of points within each Stage occurs simply because this absorption system is relatively simple, with only 2 components, such that when {\sc gvpfit} attempts to insert additional components at various trial positions within the absorbing region, the AICc value always increases and that model is thus rejected.}
\label{fig:4stages}
\end{figure}

The {\sc x-shooter} spectral resolution does not resolve individual absorbing components.  However, we simultaneously fit multiple transitions at the same redshift, with tied parameter constraints, such that the Voigt profile parameters and $\alpha$ measurements are reasonably well constrained. Nevertheless, the lower spectral resolution of {\sc x-shooter} (compared to echelle spectrographs such as {\sc uves} and {\sc hires}) would lead us to expect that some absorption components are missed. In a small number of cases, elevated $b$-parameters in the final models reinforce that expectation (full model parameter details and estimated uncertainties for all four absorption systems are provided via the online Supplementary Materials associated with this paper). Even so, Figure \ref{fig:4stages} indicates that $\alpha$ is likely to be insensitive to missing components because $\alpha$ stabilises in relatively early model generations and subsequently varies only slightly as model complexity increases.  The same insensitivity of $\alpha$ to missing components was borne out in the numerically simulated spectral simulations described in \cite{Bainbridge2017}. The impact of subsequent higher spectral resolution would evidently reduce the $\daa$ error bar but the {\sc x-shooter} results presented here should not be systematically biased by the lower resolution.

The fine structure constant has been measured in 3 high redshift absorption systems using an {\sc x-shooter} spectrum of the $z_{em}=7.084$ quasar J1120+0641.  These 3 measurements are the highest redshift direct measurements of $\alpha$ to date. The final results are summarised in Table \ref{tab:results}, giving both statistical and systematic parameter uncertainties. The weighted mean value of $\alpha$ is consistent with the terrestrial value and is $\daa = -2.18 \pm 7.27 \times 10^{-5}$. 

To update the parameters associated with the spatial dipole discussed in \cite{2011PhRvL.107s1101W, King12}, we form a new combined sample of $\alpha$ measurements as follows:
\begin{enumerate}[itemsep=-1mm]
\item The 4 new {\sc x-shooter} measurements from this paper;
\item The large sample of 293 measurements from \cite{King12};
\item 20 measurements from \cite{Wil2015}, 14 of which were re-measurements of points already in \cite{King12}, the points from \cite{Wil2015} taking priority;
\item 21 recent measurements as compiled in \cite{mp2017}.
\end{enumerate}

We opt only to use {\it direct} $\alpha$ measurements and not combined measurements (like $\alpha^2 \mu$ and $\alpha^2 g_p/ \mu$) to avoid model-dependant coupling constant assumptions. Altogether our final sample comprises a total of 323 measurements spanning the redshift range $0.2 < z_{abs} < 7.1$ enabling an updated estimate of the spatial dipole model reported in \cite{King12}: the updated dipole amplitude, \textit{A} = $0.70 \pm  0.16 \times 10^{-5}$, the dipole sky location is right ascension $17.12 \pm 0.95$ hours and declination $-57.18 \pm 8.60$ degrees. Using the bootstrap method described in \cite{King12} to estimate statistical significance, this deviates from a null result at a level of $\sim$ 3.68$\sigma$. We can also directly compare the dipole model prediction (using the new parameters above) with the actual weighted mean from the 3 new {\sc x-shooter} measurements: the dipole prediction for the weighted mean is $\daa = 0.07 \times 10^{-5}$, in agreement with the actual measurement of $\daa = -2.18 \pm 7.27 \times 10^{-5}$.

The {\sc x-shooter} data presented here highlight an important benefit that is generally not available with higher resolution echelle spectra of quasars: the extended wavelength coverage increases our ability to detect absorption systems along the line of sight simply because more transitions at the same redshift appear.  Systems that might otherwise remain undiscovered or uncertain become clear.  This is important because potential blends in transitions of interest are revealed (Table \ref{tab:identifications}) and hence systematic effects on the measurement of $\alpha$ reduced. A second advantage of the extended wavelength coverage is that since there are more transitions falling within the observed spectral range, a more stringent constraint on $\daa$ is achieved. Ultimately the precision of the very high redshift  measurements reported in this paper will be improved by obtaining higher spectral resolution, using new instrumentation such as {\sc HIRES} on the {\sc ELT}.

\bibliography{scibib}
\bibliographystyle{Science}

\section*{Acknowledgments}
Results are based on observations collected at the European Southern Observatory, Chile, programmes 286.A-5025(A), 089.A-0814(A), and 093.A-0707(A). We are grateful for the award of computing time for this research on the gStar and OzStar supercomputing facilities. MRW acknowledges support from an Australian Postgraduate Award. JKW thanks the John Templeton Foundation, the Department of Applied Mathematics and Theoretical Physics and the Institute of Astronomy at Cambridge University for hospitality and support, and Clare Hall for a Visiting Fellowship. The work of ACL and CJM was financed by FEDER---Fundo Europeu de Desenvolvimento Regional funds through the COMPETE 2020---Operational Programme for Competitiveness and Internationalisation (POCI), and by Portuguese funds through FCT---Funda\c c\~ao para a Ci\^encia e a Tecnologia in the framework of the project POCI-01-0145-FEDER-028987. ACL is supported by an FCT fellowship (SFRH/BD/113746/2015), under the FCT Doctoral Program PhD::SPACE (PD/00040/2012). We thank Julian King for useful discussions. JDB thanks the STFC for support.

\clearpage

\clearpage

\clearpage

\begin{figure}
\centering
\includegraphics[width=16cm]{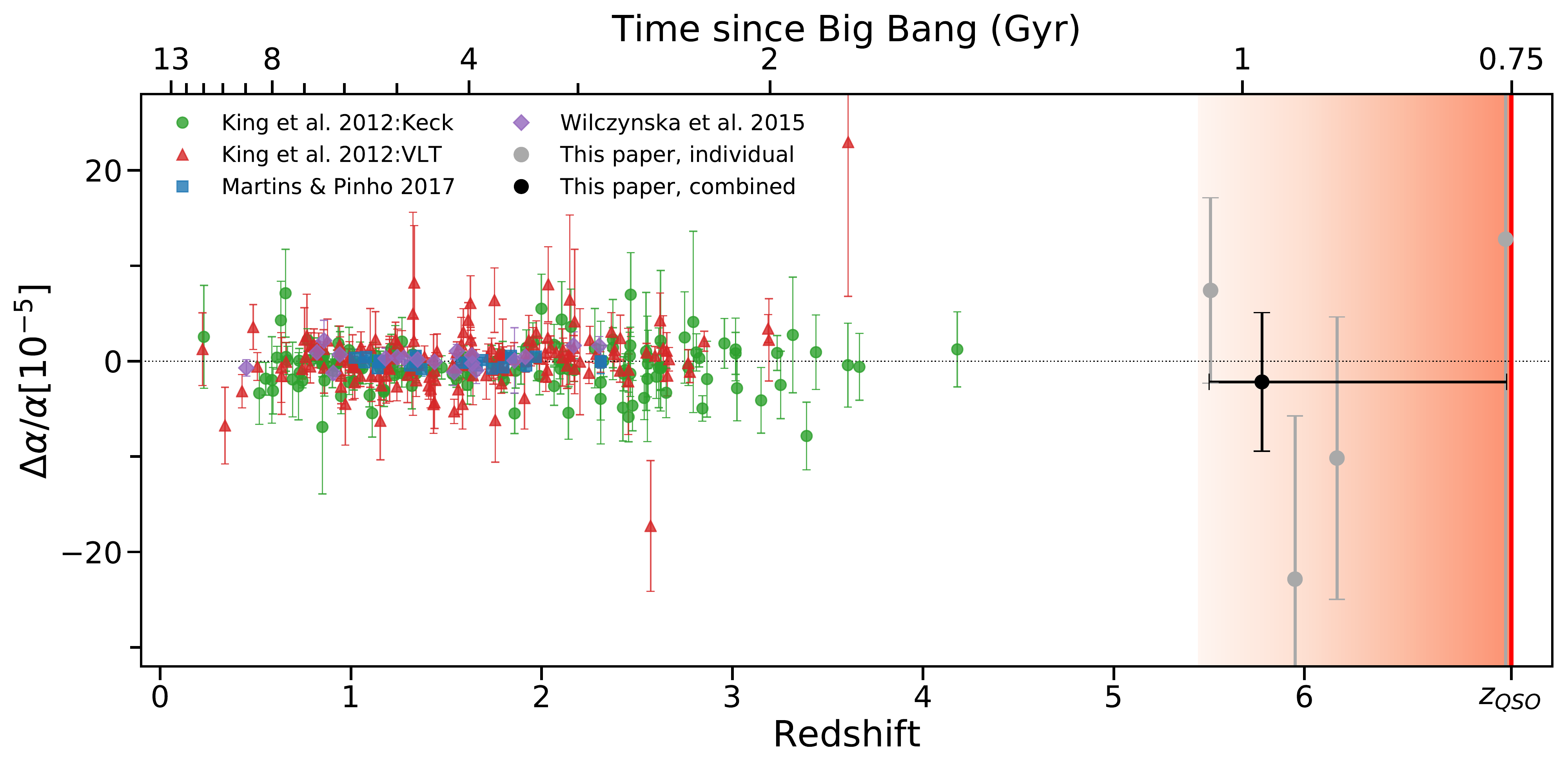} 
\caption{Direct measurements of $\daa$, taken from quasar absorption measurements \cite{King12, mp2017,Wil2015}. Where measurements reported in \cite{Wil2015} were re-analyses of the same systems from \cite{King12}, the former were used. Error bars include systematic contributions (although we note the heterogeneous nature of this combined dataset and point out that systematic errors were not all estimated in a consistent manner so error bars are not necessarily directly comparable in all cases). The point in black at $z=5.87$ illustrates the weighted mean of the 3 measurements described in this paper. Its horizontal bar indicates the redshift range spanned by those 4 measurements. The red shaded area shows the redshift range from the quasar emission redshift ($z_{em}=7.085$) down to the lowest possible redshift for a $\daa$ measurement ($z_{abs}=5.443$) assuming we retain the lowest rest-wavelength anchor line, SiII 1526 {\AA}.}
\label{fig:linearinz}
\end{figure}

\clearpage

\begin{table}[h]
\small
\centering
\begin{tabular}{lrrrrr}
\hline 
 & \multicolumn{1}{c}{Thermal} & \multicolumn{1}{c}{$\chi^2_{\nu}$} & \multicolumn{1}{c}{Turbulent} & \multicolumn{1}{c}{$\chi^2_{\nu}$} & \multicolumn{1}{c}{Distortion Corrected}\tabularnewline
$z_{abs}$ & $\Delta\alpha/\alpha\left[10^{-5}\right]$ &   & $\Delta\alpha/\alpha\left[10^{-5}\right]$ &  & $\Delta\alpha/\alpha\left[10^{-5}\right]$\tabularnewline
\hline 
7.05852 & 16.18$\pm$48.99 & 1.32 & -9.38$\pm$48.71 & 1.35 & 12.79$\pm$48.66$\pm$19.74\tabularnewline
6.17097 & -10.14$\pm$14.79 & 1.28 & -10.43$\pm$14.91 & 1.28 & -10.16$\pm$14.80$\pm$0.42\tabularnewline
5.95074 & -23.00$\pm$17.10 & 0.98 & -20.61$\pm$16.90 & 0.95 & -22.85$\pm$17.11$\pm$0.32\tabularnewline
5.50726 & 7.60$\pm$9.58 & 1.17 & 4.83$\pm$8.92 & 1.20 & 7.42$\pm$9.60$\pm$1.52\tabularnewline
\hline 
Weighted mean & 1.84$\pm$7.20\,\,\,&& -2.97$\pm$6.90 && -2.18$\pm$7.27
\,\,\,\,\,\,\,\,\,\,\,\,\,\,\,\,\,\,\tabularnewline
\hline
\vspace{0.1in}
\end{tabular}
\caption{Summary of final results. The $\daa$ uncertainties in the thermal and turbulent columns are derived from the covariance matrix (inverse Hessian) diagonal terms. The final column combines the thermal and turbulent values using the method of moments and also includes an estimated systematic term associated with possible long-range wavelength distortions (although no evidence for wavelength distortion was found). Note that even if present, its contribution to the overall $\daa$ error budget is small.  This is because of the large wavelength coverage of {\sc x-shooter} and because the distortion model parameters were derived from a simultaneous fit to all transitions in all 3 absorption systems.}
\label{tab:results}
\end{table}

\clearpage

\section*{Supplementary materials}
\subsection*{Observations and data reduction}
Observations of the quasar J1120+0641 were obtained using the {\sc x-shooter} spectrograph on the European Observatory's Very Large Telescope (VLT). The total integration time was 30 hours spanning a period from March 2011 to April 2014\footnote{ESO programmes 286.A-5025(A), 089.A-0814(A), and 093.A-0707(A).} All exposures were taken with slit widths of 0.9 arcseconds for the visual ({\sc vis}) and near infra-red ({\sc nir}) arms of the {\sc x-shooter} spectrograph, giving spectral resolutions of $R = \lambda/\Delta\lambda = 7,450$ and $5,300$ respectively. However, inspection of telluric absorption lines indicate a higher spectral resolution, suggesting the atmospheric seeing was better than the slit width used. Atmospheric absorption lines were measured as having a $R$ of $\simeq$10,000 for the {\sc vis} arm and R$\simeq$7,000 for the {\sc nir} arm, consistent with a seeing FWHM $\simeq$ 0.7 arcseconds. These values are consistent with the more detailed discussion about {\sc x-shooter} resolution given in \cite{Selsing2018}. The total wavelength coverage is approximately $5,505 - 22,740$\AA.  The spectral signal-to-noise varies across the spectrum and is approximately 21 per 10 km s$^{-1}$ pixel at $11,191$\AA.

Data reduction was performed using custom IDL routines\footnote{http://www.exelisvis.com}. The procedures include flat fielding the exposures and sky subtraction using the optimal extraction method as described by \cite{2003PASP..115..688K}. The extracted one-dimensional spectra, re-binned to 10 km s$^{-1}$ pixels, were flux calibrated using response curves derived from standard stars. Absolute flux calibration was performed by scaling the corrected spectrum to match the VLT/FORS2 and GNS spectra of J1120$+$0641 obtained by Mortlock et al. in \cite{2011Natur.474..616M}. Atmospheric line removal was performed using {\sc SkyCalc} atmospheric transmission models\footnote{http://www.eso.org/sci/software/pipelines/skytools/}. A comprehensive description of the observations and data reduction is given in \cite{Bos17}.

\subsection*{Continuum fitting}
Prior to profile fitting the absorption systems of interest, we need a reliable estimate of the unabsorbed quasar continuum.  This was obtained using the {\sc iraf}\footnote{http://iraf.noao.edu/} task {\sc continuum}.  Small spectral regions flanking each absorption line were selected and the pixels containing absorption were masked. The spectral regions used to estimate the underlying continuum (by fitting cubic splines, typically of order 3) each contained $\sim 100$ to $300$ pixels. 

\subsection*{Identification of absorption systems}
Identification of absorption systems and atomic species present was carried out using {\sc qscan} \cite{2017Dum}, an interactive Python program to display the spectrum on a velocity scale such that, at fixed redshift, transitions from different absorption species align in velocity space. Absorption systems were identified by scanning through absorption redshift and searching for alignments in velocity space at each redshift. The spectral ranges chosen for fitting were selected as described in \cite{Wil2015}. Eleven absorption systems were detected (Table 1). Of these eleven, four were selected for their sensitivity to any variation of the fine-structure constant; $z_{abs}=7.05852, z_{abs}=6.17097$, $z_{abs}=5.95074$, and $z_{abs}=5.50793$. We excluded Si IV and C IV in the determination of $\alpha$ in the three latter systems as the ionization potentials of these transitions are significantly higher than the other (more sensitive) transitions available and higher ionization lines may be spatially (and hence velocity) segregated from the lower ionization transitions, potentially emulating a change in $\alpha$. MgII 2796 in absorption system at $z_{abs}=5.50794$ fall in a region of the spectrum containing an incompletely removed telluric line and was not included in the modelling. For the highest redshift system, $z_{abs}=7.059$, only high ionization species were available.  The lower sensitivity of these results in a substantially larger error on $\alpha$ but including the system has the advantage of producing a tighter constraint on any possible long-range distortion in the spectrum, hence improving the overall result.

\begin{table}[h]
\begin{tabular}{llcc}
\hline 
$z_{abs}$ & Transitions {(\AA)}\tabularnewline
\hline 
\rule{0pt}{2.5ex}\textbf{7.05852} & \textbf{C IV 1548/1550}, \textbf{Si IV 1393/1402}, \textbf{N V 1242/1238} \tabularnewline
7.01652  & C IV 1548/1550 \tabularnewline
6.51511  & C IV 1548$^{1}$/1402 \tabularnewline
6.40671  & Mg II 2796/2803 \tabularnewline
6.21845  & C IV 1548/1550, Mg II 2796/2803 \tabularnewline
\textbf{6.17097} & \textbf{Al II 1670}, C IV 1548/1550$^{1}$, \textbf{Si II 1526}, \textbf{Fe II 2383}, \tabularnewline & \textbf{Mg II 2796/2803}, Si IV 1393$^{2}$/1402 \tabularnewline
\textbf{5.95074} & \textbf{Fe II 2344/2383/2587/2600,} \textbf{Mg II 2796$^{3}$/2803$^{3}$, Si II 1526} \tabularnewline
5.79539  & CIV 1548/1550 \tabularnewline
\textbf{5.50726} & Al II 1670, \textbf{Fe II 2344/2383}/2587$^{4}$/\textbf{2600$^{5}$/1608,}  \textbf{Mg II} 2796$^{4}$/\textbf{2803, Si II 1526} \tabularnewline 4.47260  & Mg II 2796/2803 \tabularnewline
2.80961  & Mg II 2796/2803 \tabularnewline
\hline 
\multicolumn{4}{l}{$^{1}$\rule{0pt}{2.5ex} Line is contaminated by C IV 1548 from intervening absorption system at $z_{abs}=6.51511$.\newline}\tabularnewline
\multicolumn{4}{l}{$^{2}$ Line is contaminated by N V 1238 from intervening absorption system at $z_{abs}=7.05852$.\newline}\tabularnewline
\multicolumn{4}{l}{$^{3}$ Mildly affected by cosmic rays. \newline}\tabularnewline
\multicolumn{4}{l}{$^{4}$ Line is blended with incompletely removed telluric line. \newline}\tabularnewline
\multicolumn{4}{l}{$^{5}$ Broad interloper at -100 km s$^{-1}$.}
\vspace{0.1in}
\end{tabular}
\caption{Absorption systems and transitions identified in the {\sc x-shooter} spectrum of the $z_{em} = 7.084$ quasar J1120+0641. Absorption redshifts are listed in column 1. Lists of transitions present in each absorption system are listed in column 2. The 4 absorption systems and transitions used to measure $\Delta\alpha/\alpha$ are indicated in bold.}
\label{tab:identifications}
\end{table}

\subsection*{Atomic data and sensitivity coefficients}

We used multiplets from different atomic species simultaneously to constrain any possible variation of the fine structure constant $\alpha$.  The method used, the {\it Many Multiplet Method}, was introduced in \cite{1999PhRvL..82..888D, 1999PhRvL..82..884W}.  Sensitivity coefficients (q-coefficients, parameterising the sensitivity of an observed wavelength to variation of the fine-structure constant) are compiled in \cite{MB2014} along with Laboratory wavelengths, oscillator strengths, and hyperfine structure and spontaneous emission rates ($\Gamma$ values). 

\begin{figure}
\centering
\includegraphics[width=16cm]{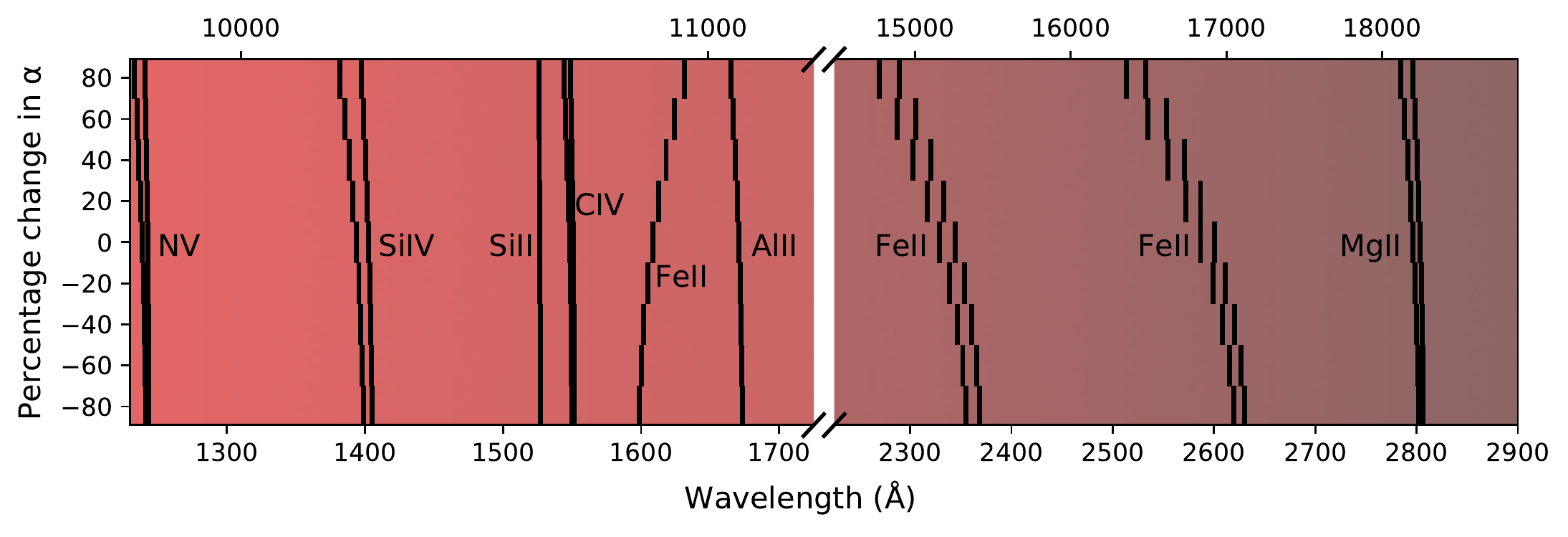}   \caption{Illustration of the $\alpha$ dependence of the transitions observed in the 4 absorption systems measured in this paper. The percentage change in $\alpha$ has been exaggerated to show the shift trends. The lower x-axis is rest-wavelength, the upper observed-frame wavelength at redshift $z=5.87$ (the mean redshift for all 3 absorption systems in this analysis). Some transitions (SiII 1526, AlII 1670, and MgII 2796/2803 {\AA}) are insensitive to changes in $\alpha$ (``anchors''), whilst the FeII transitions all show a substantially greater sensitivity, with FeII 1608 {\AA} shifting in the opposite direction to FeII 2344, 2383, 2586, and 2600 {\AA}. The sensitivity of an observed frequency $\omega_z$ at redshift $z$ to a change in $\alpha$ is given by $\omega_z = \omega_0 + q (\alpha_z^2/\alpha_0^2 - 1) \approx 2 \delta\alpha/\alpha$ where $\alpha_0$ and $\omega_0$ are the terrestrial values and $q$ is the sensitivity coefficient for that transition.}

\label{fig:tshifts}
\end{figure}

Rest-frame wavenumbers of atomic transitions observed in quasar absorption spectra are related to laboratory values via the relationship $\omega_{z}=\omega_{0}+q(\alpha_{z}^{2}-\alpha_{0}^{2})/\alpha_{0}^{2}$, where $\omega_{z}$ is the wavenumber at redshift $z$, $z=\lambda_{obs}/\lambda_{lab}-1$, $\omega_{0}$ is the laboratory value, and $q$ is a coefficient parameterizing the sensitivity of a given transition to a change in $\alpha$. The large non-ordered range in $q$-coefficients and their different signs create a unique varying $\alpha$ signature and assist in overcoming simple systematic effects. Figure \ref{fig:tshifts} shows how the transition wavelengths of SiII, AlII, FeII, and MgII (the transitions used in this analysis) depend on $\alpha$.  The range in $\alpha$ is grossly exaggerated for illustration.

\subsection*{Further details and final model adjustments}

The {\sc gvpft} modeling produces near-final fits but additional physical considerations, not coded into the AI methodology, are helpful in deriving the final absorption line models.  These relatively minor tweaks to the non-linear least-squares input-guess parameters were done (a) to remove or minimize the presence of parameters that are physically implausible and (b) to further improve the model by reducing the overall $\chi^2$. The notes here record those considerations and justify final model parameters.\\

\noindent $z_{abs}=5.507261$: \newline 
No changes to the {\sc gvpfit} models were required. The following discussion applies to both the thermal and the turbulent fit. 

The {\sc gvpfit} model for this system included an interloper at approximately $+100$ km s$^{-1}$ in FeII 2600 {\AA} (see Figure \ref{fig:therm5-50}) for which the column density and $b$-parameter were poorly constrained. Removing this interloper resulted in a negligible change to either the overall $\chi^{2}$ or to any of the remaining model parameters, so the interloper has been excluded. 

Visual inspection shows a very broad shallow continuum depression over the FeII 2600 {\AA} absorption line at approximately $-80$ km s$^{-1}$, as Figure \ref{fig:therm5-50} illustrates.  {\sc gvpfit} modelled this using a high-$b$ component ($b=81$ km s$^{-1}$) straddling the whole absorption complex.  There is no species identification for this interloper. It may be due to real unidentified absorption or it may be some observational artefact. The associated degeneracy in the final model due to these additional parameters is minimal and impacts negligibly on the other model parameters.\\

\begin{figure}
\centering
\includegraphics[width=16cm]{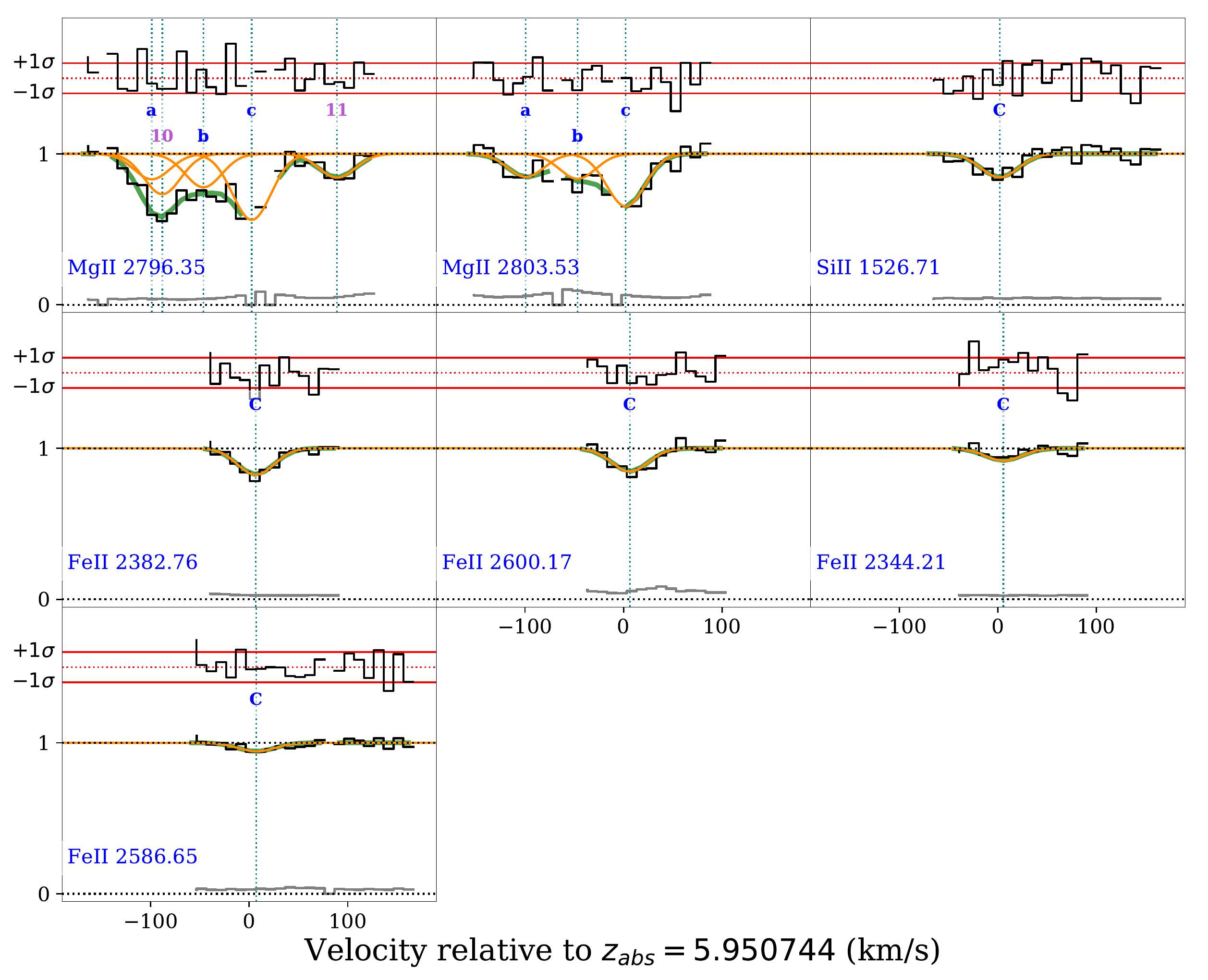}    
\caption{Thermal fit for the $z_{abs}=5.950744$ absorption system.}
\label{fig:therm5-95}
\end{figure}

\noindent $z_{abs}=5.950744$: \newline
Thermal fit: The MgII lines reveal 3 components.  Interestingly the {\sc gvpfit} model found an interloper heavily blended with the left-hand MgII component at approximately $-20$ km s$^{-1}$.  Cosmic ray events on the detector spoil the FeII lines in this region so the leftmost component provides only a very weak constraint on $\daa$ which is thus constrained almost entirely by the right hand component at $+75$ km s$^{-1}$.  Once the initial model fits were available, it became apparent that several pixels in the MgII 2796 {\AA} and 2803 {\AA} lines were significantly deviant.  We assumed this was due to weak cosmic rays contaminating the spectrum.  We thus manually clipped 3 pixels in the 2796 {\AA} line and 2 pixels in the 2803 {\AA} line, as can be seen in Figure \ref{fig:therm5-95}.\\

\noindent Turbulent fit: The independently derived turbulent model found by {\sc gvpfit} differs to the thermal model in that no interloper was assigned to the leftmost MgII component. Whilst this makes the model rather different to the thermal one, the impact on $\daa$ and its uncertainty is minimal and both thermal and turbulent models yield a consistent result for $\daa$.\\

\noindent $z_{abs}=6.170969:$ \newline
No changes to the {\sc gvpfit} model were made for this absorption system. This absorption system was modelled by {\sc gvpfit} as single component. The $b$-parameters for the transitions seen in this system are comparable with those found when modelling higher resolution data.  No interlopers were identified.\\

\noindent $z_{abs}=7.058521:$ \newline
The CIV and NV lines are strong, visually comprising  2-strong components, which are shown by careful (automated) modelling to break into further components. SiIV is weak so contributes little to the $\alpha$ constraint despite being more sensitive to a change in $\alpha$. No model changes were made to the automated fit.

\subsection*{Checking for long-range wavelength distortions}

Long-range wavelength distortions have been discovered and measured in the Keck-HIRES, VLT-UVES, and Subaru-HDS spectrographs \cite{2003PASP..115.1050S}, \cite{RWS2013}, \cite{EMW14}, \cite{MW15}.  Whilst no analogous distortions have been identified in {\sc x-shooter}, in this analysis we apply caution and assume that they could be present.  It has recently been shown that such distortions can be modeled independently of any additional calibration exposures \cite{DW17}.  We use the same method used in \cite{DW17} to model a putative distortion, taking advantage of the presence of multiple absorption systems along the same line of sight.  The presence of many transitions spread over a wide range in observed wavelength allows us to place tight constraints on any possible distortion.

Previous studies have found that the functional form of the long-range distortions (found in solar-twin and asteroid measurements) are approximately linear, with no shifts found at the central wavelength of the science exposure \cite{RWS2013,MW15}. That is, they can be parameterized as a linear fit, with slope $\gamma$ (m/s/\AA), passing through the central wavelength of the exposure. We have adopted these assumptions for determining a best-fit distortion model for J1120+0641.

\begin{figure}
\centering
\includegraphics[width=0.9\linewidth]{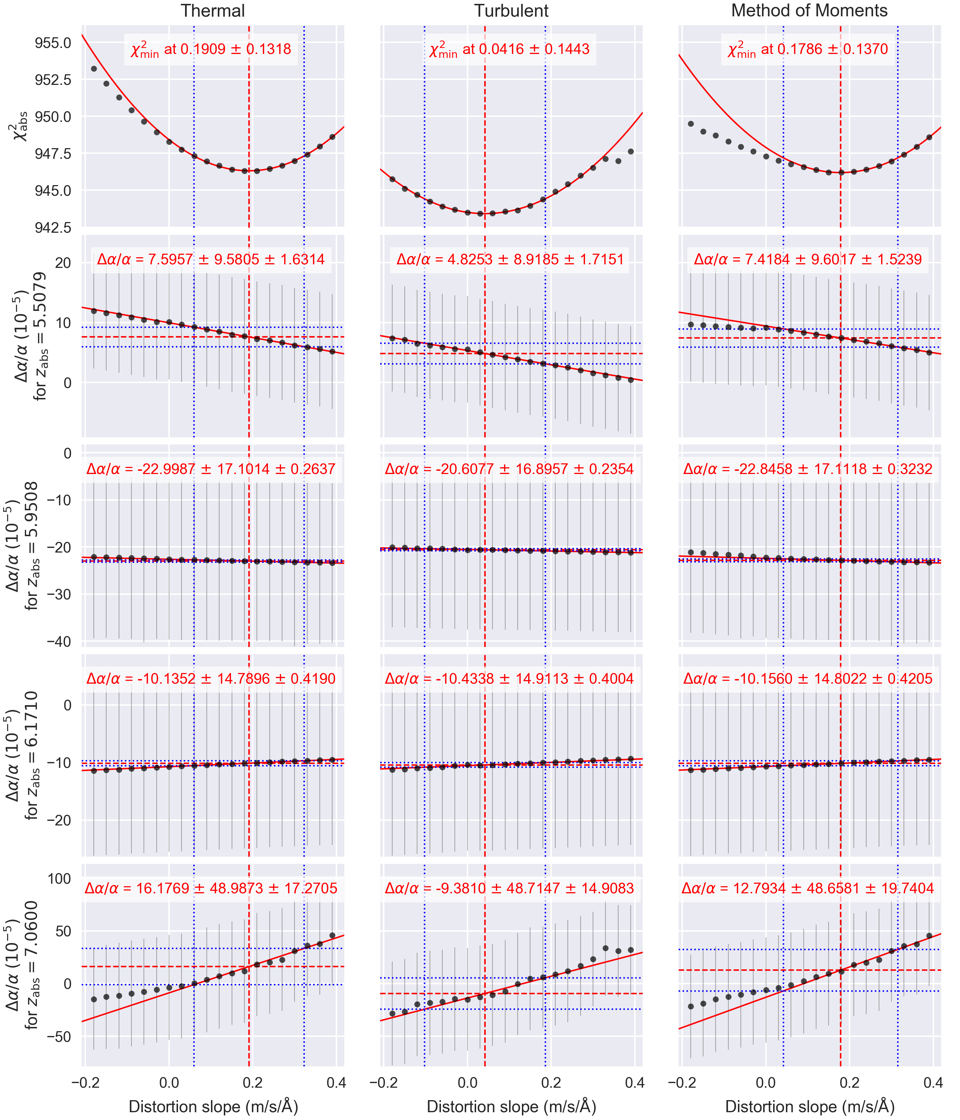}
\caption{Distortion analysis for J1120+0641 using transitions from all 4 absorption systems simultaneously to derive the best-fit linear distortion model. Top row: overall $\chi^2$ showing best-fit slopes for thermal and turbulent models and the combined result using method of moments; Subsequent rows: estimated impact on $\daa$ for each absorption system. Vertical error bars (thin grey lines): the {\sc vpfit} uncertainty on $\daa$. Red dashed vertical lines: the best-fit slope. Blue dotted vertical lines: the $\pm 1\sigma$ ranges on the distortion slope. Blue dotted horizontal lines: the inferred additional systematic error on $\daa$ given the best-fit slope and its uncertainty. The uncertainty on $\daa$ due to distortion is small compared to the {\sc vpfit} uncertainty.}
\label{fig:distortion}
\end{figure}

We solve externally for the slope of a simple distortion model.  The slope is varied in steps of $\delta\gamma=0.05$ m/s/{\AA} in the range $-1.0$ $\leq \gamma \leq 1.0$ m/s/{\AA} and {\sc vpfit} is used to solve for the absorption model parameters at each step.  The overall $\chi^2$ is then minimized as a function of distortion slope, as described in \cite{DW17}. Using all four absorption systems simultaneously to solve for the distortion slope, we find  $\gamma = 0.15 \pm 0.19$ m/s/{\AA} (Figure \ref{fig:distortion}). We thus find no significant evidence for long-range distortion.  We nevertheless include an additional term in the final error $\daa$ budget corresponding to this $\gamma$ value. 

\clearpage

\begin{figure}
\label{therm5-50}
\centering
\includegraphics[width=16cm]{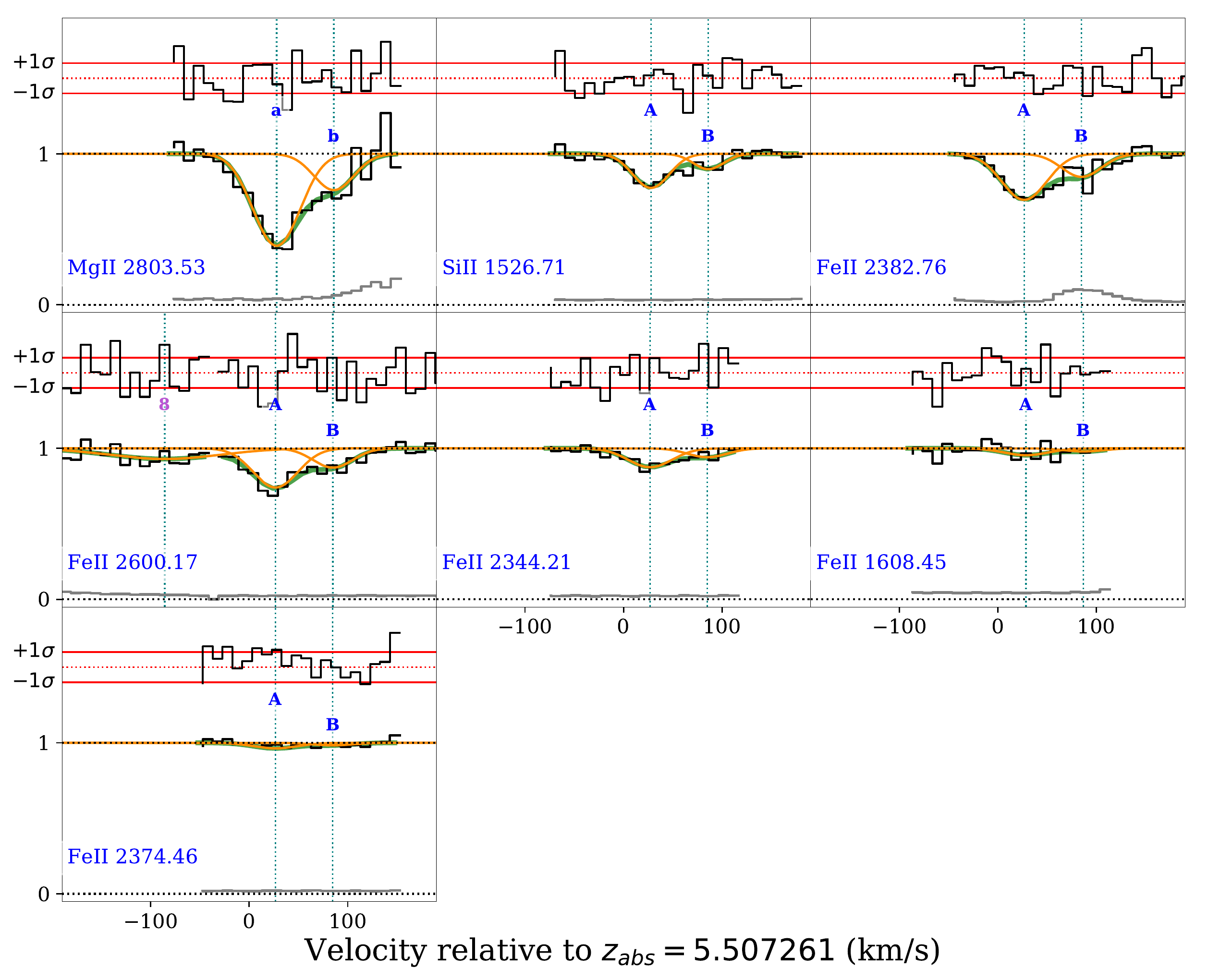}    
\caption{Turbulent fit for the $z_{abs}=5.507261$ absorption system.}
\label{fig:turbm5-50}
\end{figure}

\begin{figure}
\centering
\includegraphics[width=16cm]{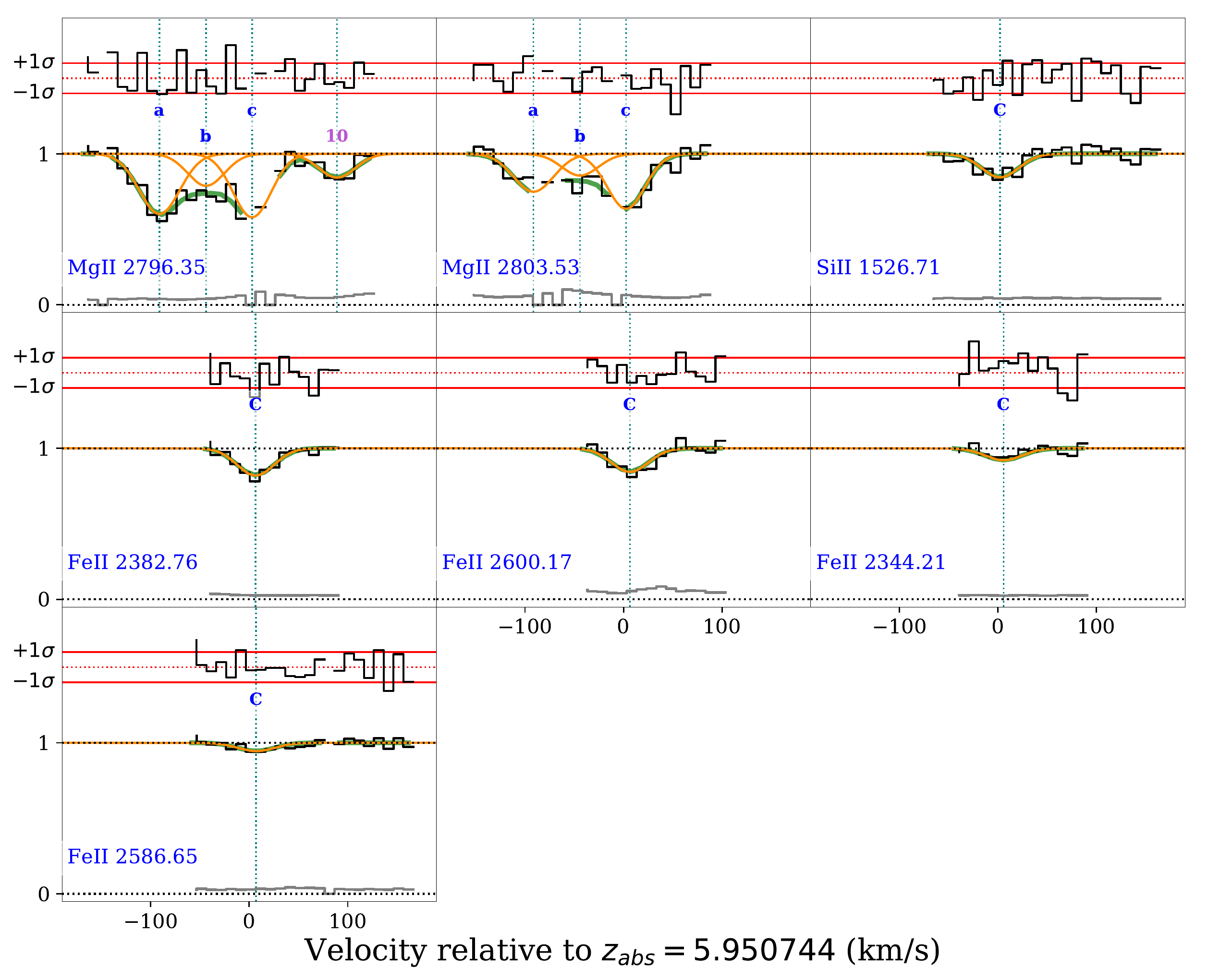}    
\caption{Turbulent fit for the $z_{abs}=5.950744$ absorption system.}
\label{fig:turb5-95}
\end{figure}

\begin{figure}
\centering
\includegraphics[width=16cm]{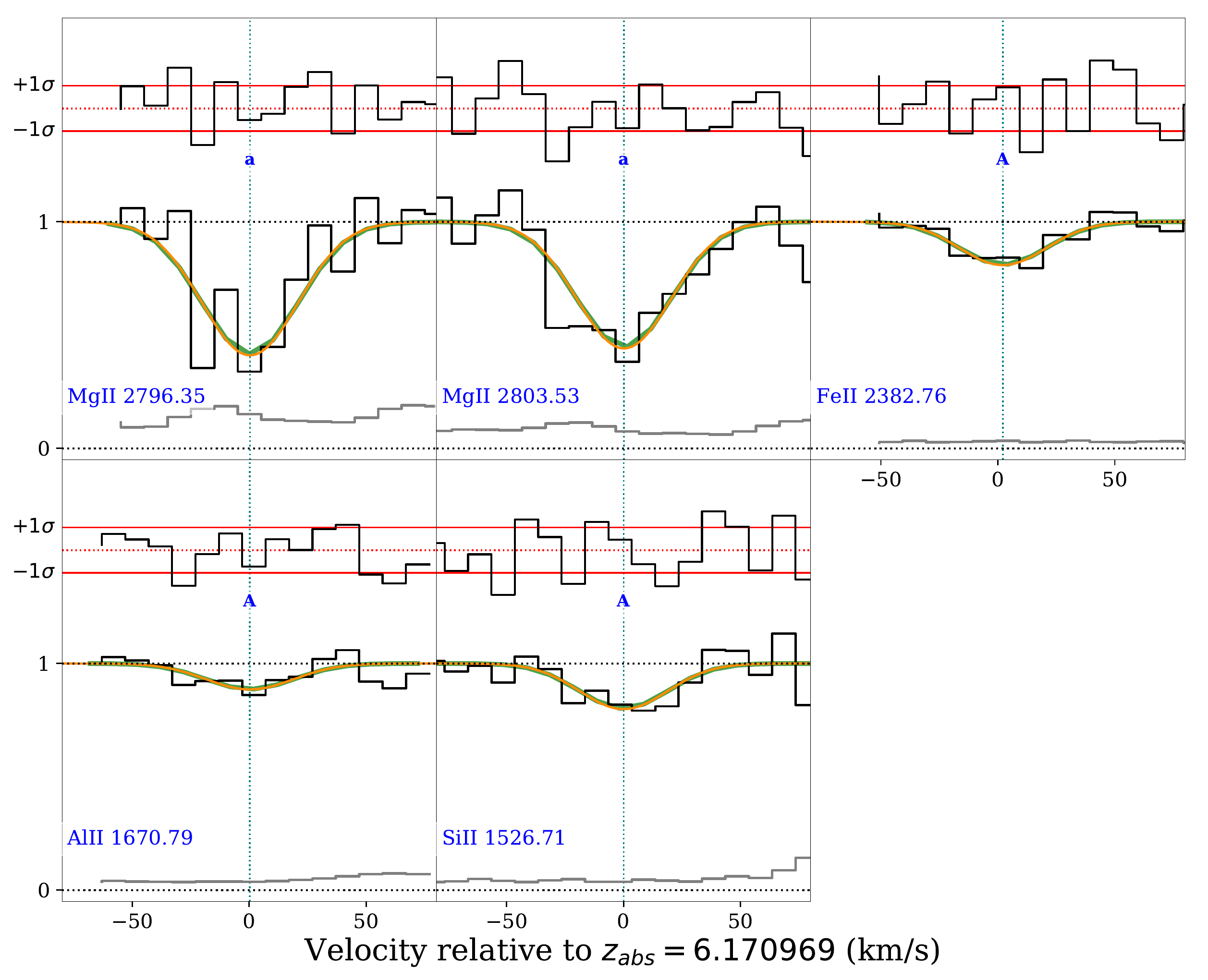}    
\caption{Thermal fit for the $z_{abs}=6.170969$ absorption system.}
\label{fig:therm6-17}
\end{figure}

\begin{figure}
\centering
\includegraphics[width=16cm]{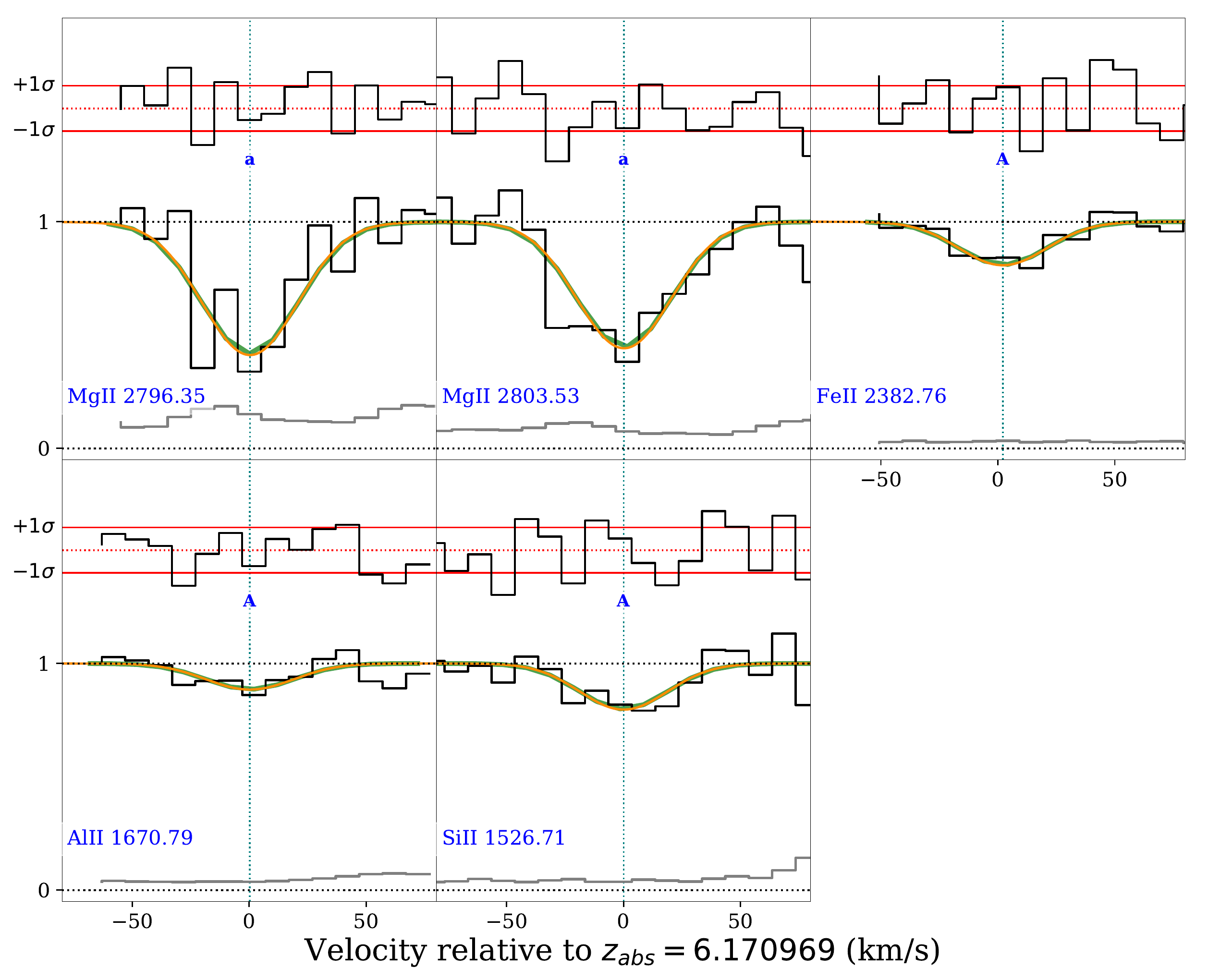}    
\caption{Turbulent fit for the $z_{abs}=6.170969$ absorption system.}
\label{fig:turb6-17}
\end{figure}

\begin{figure}
\centering
\includegraphics[width=16cm]{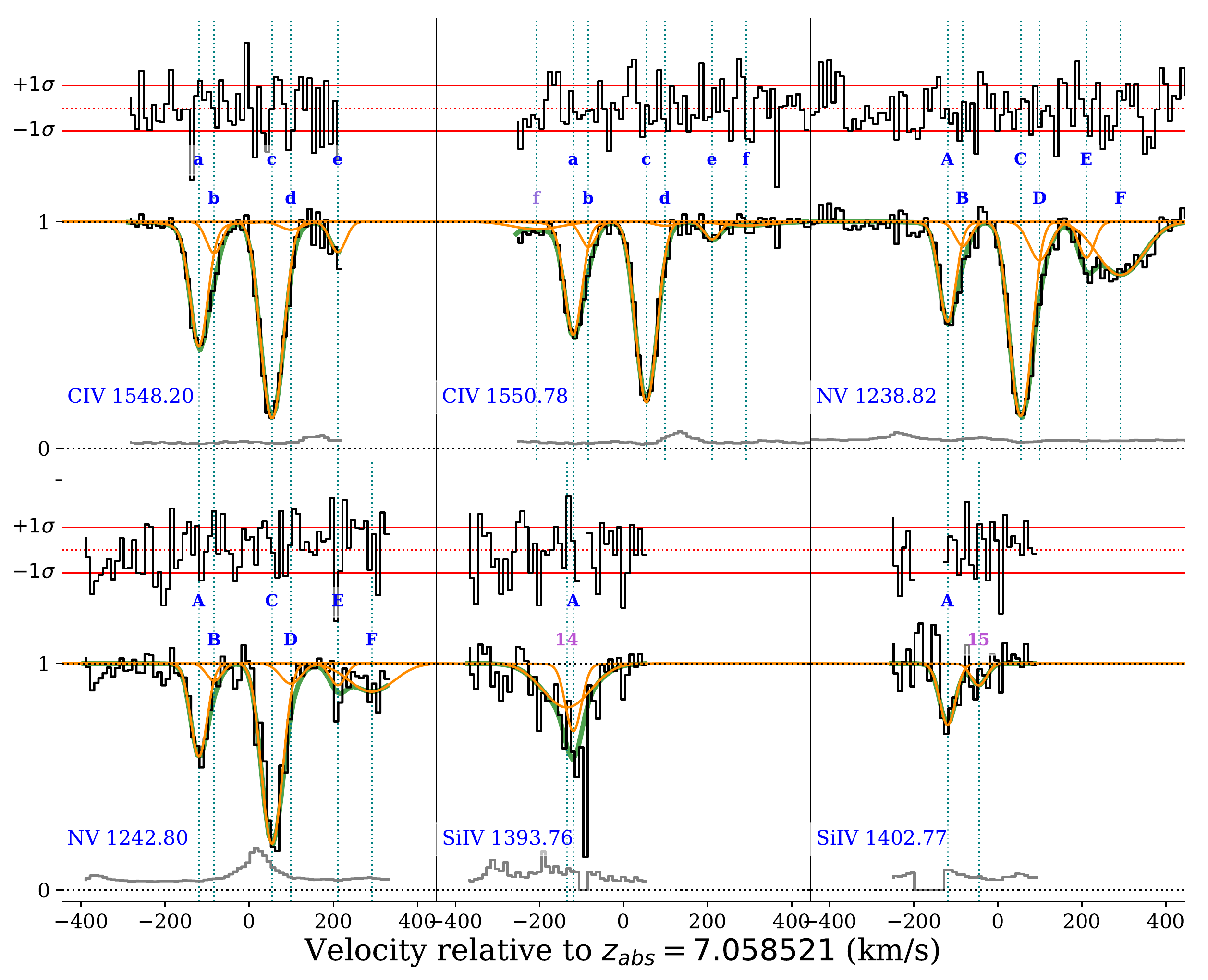}    
\caption{Thermal fit for the $z_{abs}=7.058521$ absorption system.}
\label{fig:therm7-06}
\end{figure}

\begin{figure}
\centering
\includegraphics[width=16cm]{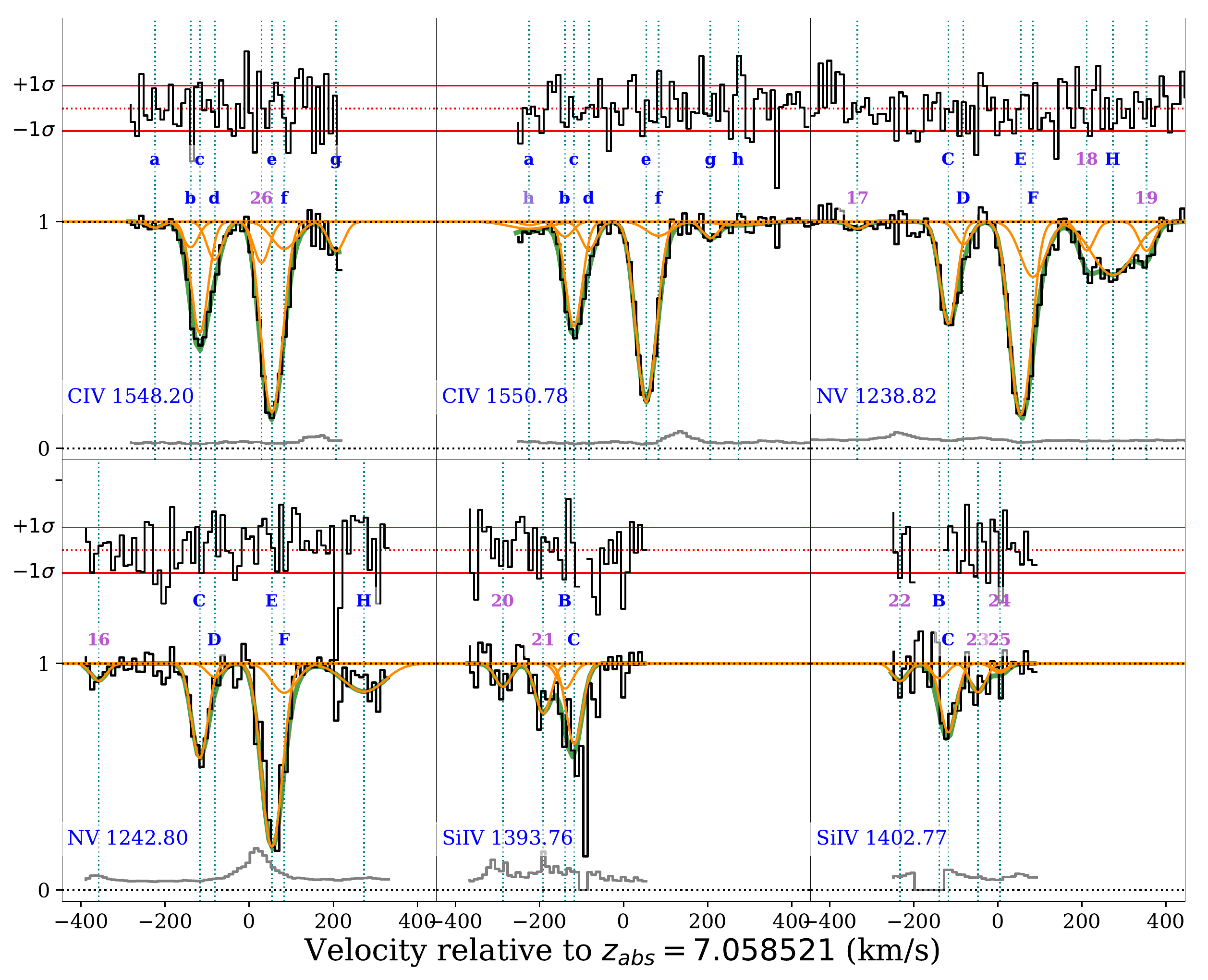}    
\caption{Turbulent fit for the $z_{abs}=7.058521$ absorption system.}
\label{fig:turb67-06}
\end{figure}

\clearpage

\end{document}